\documentclass[aps,prb,showpacs,twocolumn,superscriptaddress]{revtex4-1}

\usepackage{graphicx}
\usepackage{color}
\usepackage{enumerate}
\usepackage{amsmath}
\usepackage{amssymb}
\usepackage{stmaryrd}
\usepackage{multirow}
\usepackage{float}
\usepackage{longtable,booktabs}
\usepackage{hyperref}
\usepackage{physics}
\usepackage{soul} %to cross the text
\usepackage[colorinlistoftodos]{todonotes}
\usepackage[draft]{pdfcomment}

\newcommand{\D}{\mathrm{d}}

\def\beq{\begin{equation}}
\def\eeq{\end{equation}}

\begin{document}

\title{Bulk Geometry of the Many Body Localized Phase from Wilson-Wegner Flow}

\author{Xiongjie Yu}
\affiliation{Department of Physics, University of Illinois at Urbana-Champaign, IL 61801, USA} 

\author{David Pekker}
\affiliation{Pittsburgh Quantum Institute and Department of Physics and Astronomy, University of Pittsburgh, PA 15260, USA}

\author{Bryan K.  Clark}
\affiliation{Department of Physics, University of Illinois at Urbana-Champaign, IL 61801, USA} 

\begin{abstract}
Tensor networks are a powerful formalism for transforming one set of degrees of freedom to another.  They have been heavily used in analyzing the geometry of bulk/boundary correspondence in conformal field theories.  Here we develop a tensor-network version of the Wilson-Wegner Renormalization Group Flow equations to efficiently generate a unitary tensor network which diagonalizes many-body localized Hamiltonians.  Treating this unitary tensor network as a bulk geometry, we find this emergent geometry corresponds to the shredded horizon picture: the circumference of the network shrinks exponentially with distance into the bulk, with spatially distant points being largely disconnected. 

\end{abstract}
 
\maketitle	
Unitary tensor networks (UTN) can be used to efficiently represent strings of quantum operators. Both in the case of many-body localization (MBL) and holography, UTN have been used to transform between two types of complementary descriptions of the physical system.  

In the case of holography, UTN implement the bulk/boundary correspondence mapping boundary states to bulk states.  This idea was partially inspired by the fact that both the ``vertical'' direction in a MERA tensor network and the radial direction of the holographic bulk can be thought of as a renormalization flow~\cite{swingle2012}. Tensor network models of holography capture many important aspects of the holographic correspondence including obeying the Ryu-Takayanagi formula~\cite{Ryu} and mapping bulk to boundary operators in a redundant fashion~\cite{tHooft1993Dimensional,Susskind1995hologram,Maldacena1999Superconformal,Witten1998AdS,AHARONY2000LargeN,Swingle2012PRBEntanglement,Evenbly2011Tensor,Nozaki2012,Qi2013arXivExact,Beny2013Causal,Mollabashi2014,Pastawski2015Holographic,Bao2015Consistency,Miyaji2015Surface,Czech2016Tensor,Hayden2016Holographic,You2016Entanglement,Yang2016Bidirectional,Qi2017Holographic,Hyatt2017arXivExtracting}.  Geometry of entanglement has also been investigated using quantum circuits. \cite{hyatt2017extracting,You2017arXivMachine}

In the case of many-body localization~\cite{Basko2006,Pal2010}, the UTN transforms a set of l-bits, $n$ commuting Hermitian operators $\tau^z_i$, to p-bits, the physical degrees of freedom $\sigma^\alpha$, via  $\tau^z = U \sigma^z U^\dagger$~\cite{Huse2014,Chandran2015,Pekker2014a,Chandran2014,Pekker2016Fixed}.   
The same UTN transforms from the original Hamiltonian to the diagonal $l$-bit Hamiltonian $H= \sum_i J_i \tau^z_{i} +  \sum_{i,j} J_{ij} \tau^z_{i}\tau^z_{j} + ...$. 
The commuting operators are responsible for the emergent integrability which drives the phenomenology of the many-body localized phase including its failure to thermalize and conduct~\cite{Basko2007}; the area-law entanglement and Poisson spectral statistics of the eigenstates~\cite{Pal2010,Bauer2013}; and the slow buildup of entanglement under dynamics~\cite{Bardarson2012}. Furthermore, at the MBL transition interactions between l-bits acquire a scale invariant form~\cite{Pekker2016Fixed}. In the MBL phase, it has been shown~\cite{Pekker2014a} that the bond-dimension of these unitary tensor networks grows slowly with system size.  This led to the suggestion that these UTNs could be used to variationally diagonalize the entire MBL spectrum and this program has been partially carried out in ref.~\onlinecite{Pollmann2016} and \onlinecite{Wahl2016}.

\begin{figure}[th]
\centering
	\includegraphics[width=\columnwidth]{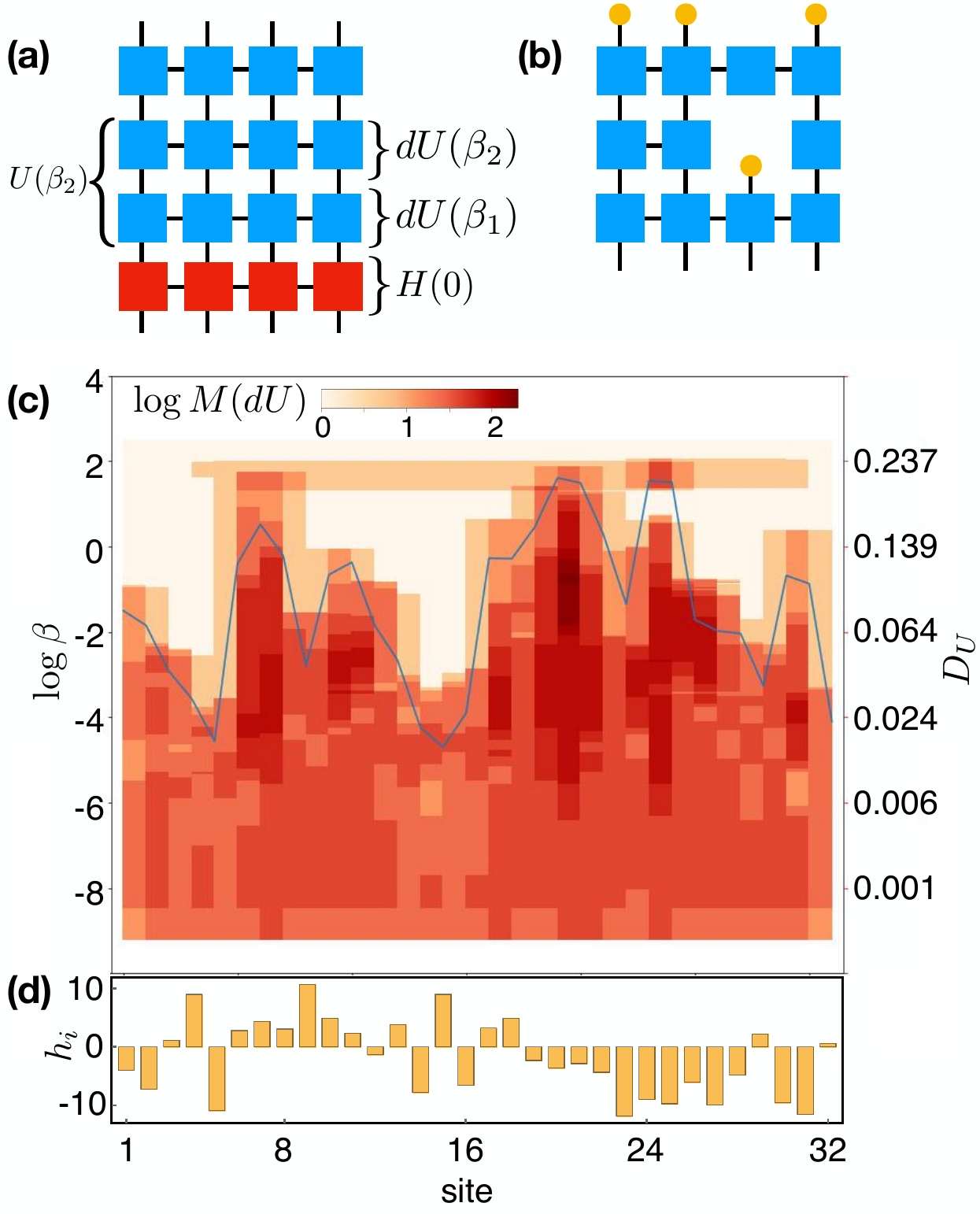}
\caption{\textbf{(a)} Hamiltonian as a matrix product operator [red] being acted on by a unitary tensor network (UTN) [blue]. The UTN consists of a stack of infinitesimal unitary transformations, represented as matrix-product operators, $dU(\beta_i)$ at time $\beta_1, \beta_2, \ldots$. The composition of these infinitesimal unitary transformations is $U(\beta)$.  \textbf{(b)} The UTN can be transformed into a tensor network for an eigenstate
by applying a product state (yellow dots) to the top.  After the application, if there are states which aren't being rotated by a unitary, the vertical bonds can be pulled down to smaller $\beta.$  \textbf{(c)} Log of bond-dimension of $dU(\beta)$ of prototypical UTN at $L=32$ and $W=12$ for the disorder distribution in \textbf{(d)}.  The blue line indicates where the coupling constants of $H(\beta)$, that are anchored to the particular site, stop changing, defined as changing at less then $1\%$ of their maximum rate of change. }
\label{fig:UTN}
\end{figure}

\begin{figure}[th]
\centering
\includegraphics[width=\columnwidth]{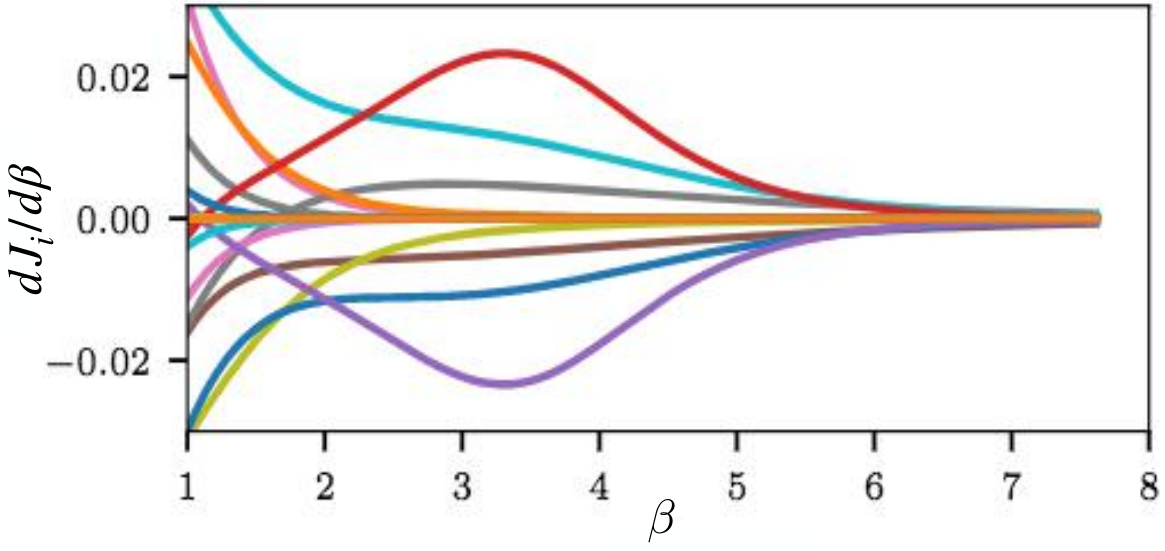}
\caption{Rate of change of the on-site couplings $dJ_i/d\beta$ plotted as a function of $1<\beta<8$ for sites $i=1,2,\dots,32$ for the sample in Fig.~\ref{fig:UTN}(d).
\label{fig:spaghetti}
%L=32, W=12, RS=3 sample. heat map with $S^z$ operators' coefficients.		
}
\end{figure}

\begin{figure}[t]
\centering
\includegraphics[width=\columnwidth]{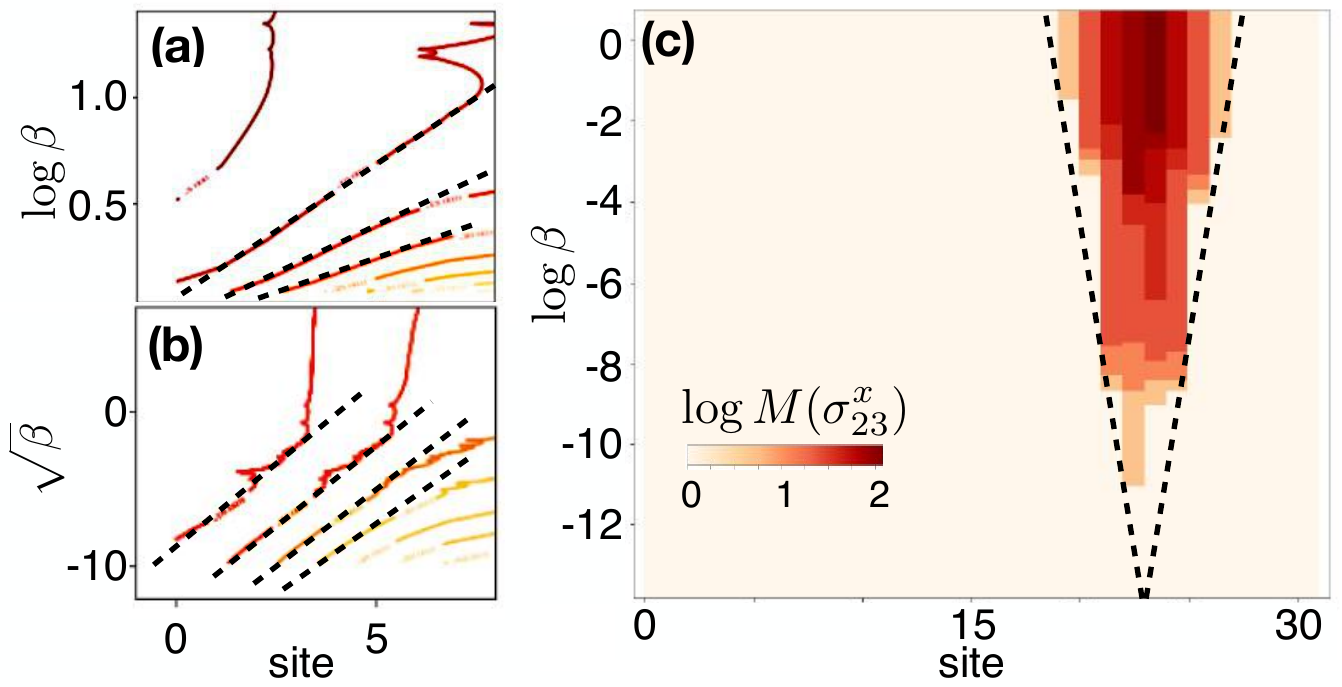}
\caption{Spread of operators through the WWF.  \textbf{(a,b)} Contours of fixed $\ln(\tr(\sigma_0^x \sigma_0^x))$ for prototypical runs at
L=8 \textbf{(a)} W=12 and \textbf{(b)} W=0.5. Dashed lines are guides to the eye. \textbf{(c)} Log of bond-dimension
of $U(\beta)\sigma_{23}^x U(\beta)^\dagger$ for the UTN $U(\beta)$ in \ref{fig:UTN}(c) as a function of $\log(\beta)$.  
%L=32, W=12, RS=3 sample. heat map with $S^z$ operators' coefficients.		
}
\label{fig:lightCone}
\end{figure}

In this work, we developed a numerical method, the Tensor Wilson Wegner Flow (TWWF), to generate a UTN which induces a holographic bulk/boundary correspondence between the boundary MBL Hamiltonian and the tensors in the bulk.  Specifically, we have developed an efficient tensor-network implementation of the Wegner-Wilson flow equations that we use to fully diagonalize MBL spin chains of up to 32 spins (TWWF is used throughout this work for $L=\{16,32\}$ while standard ED WWF is used for $L\leq 8)$. The connection to the holographic principle is made manifest by having the transverse direction of our UTN corresponds to a renormalization flow; in our case, different levels of the UTN correspond to diagonalizing the Hamiltonian up to different fixed energy scales.  We probe the bulk geometry by measuring the properties of the tensors which make up the UTN (Figs.~\ref{fig:UTN} \& \ref{fig:spaghetti}) and the rate at which operators propagate through the bulk degrees of freedom (Fig.~\ref{fig:lightCone}). We find that the circumference of the bulk shrinks exponentially in the transverse direction.  

We apply TWWF to the disordered Heisenberg model~\cite{Oganesyan2007,Pal2010,Luitz2015,Devakul2015,Khemani2016,Yu2015,Luitz2016,Luitz2016a,Yu2016,Lim2016,Znidaric2008,Luca2013,Pekker2014a,Serbyn2015,Bera2015,Luitz2016b,Pollmann2015,Singh2016,Wahl2016}
\begin{align}
\label{eq:model}
H &= \sum_i \left( \vec{S}_i \cdot \vec{S}_{i+1} + h_i S^z_i \right), \\
h_i&\in[-W,W], \quad p(h_i)=1/(2W).
\end{align}
This Hamiltonian is known to have an ergodic phase at $W \lesssim 4$ and a many-body localized phase at $W \gtrsim 4$ allowing us to probe the bulk degrees of freedom of both phases.  

The Wegner-Wilson flow (WWF)~\cite{Wegner2001,Quito2016,Pekker2016Fixed,Monthus2016} equations  are
\begin{equation}\label{eq:Flow_Equation}
\frac{\D H(\beta)}{\D \beta} = [\eta(\beta), H(\beta)], \quad \frac{\D U(\beta)}{\D \beta} = \eta(\beta) U(\beta).
\end{equation}                               
where 
\begin{equation}\label{eq:Generator}
\eta(\beta) = [H_0(\beta), H_1(\beta)],
\end{equation}
$H_0(\beta)$ and $H_1(\beta)$ are respectively the diagonal and off-diagonal parts of the Hamiltonian, and $H(0)$ is the original Hamiltonian. In the MBL problem, WWF has been shown to be a good heuristic for constructing maximally local l-bits~\cite{Pekker2016Fixed,kelly2019exploring}.

\begin{figure}[th]
\centering
\includegraphics[width=\columnwidth]{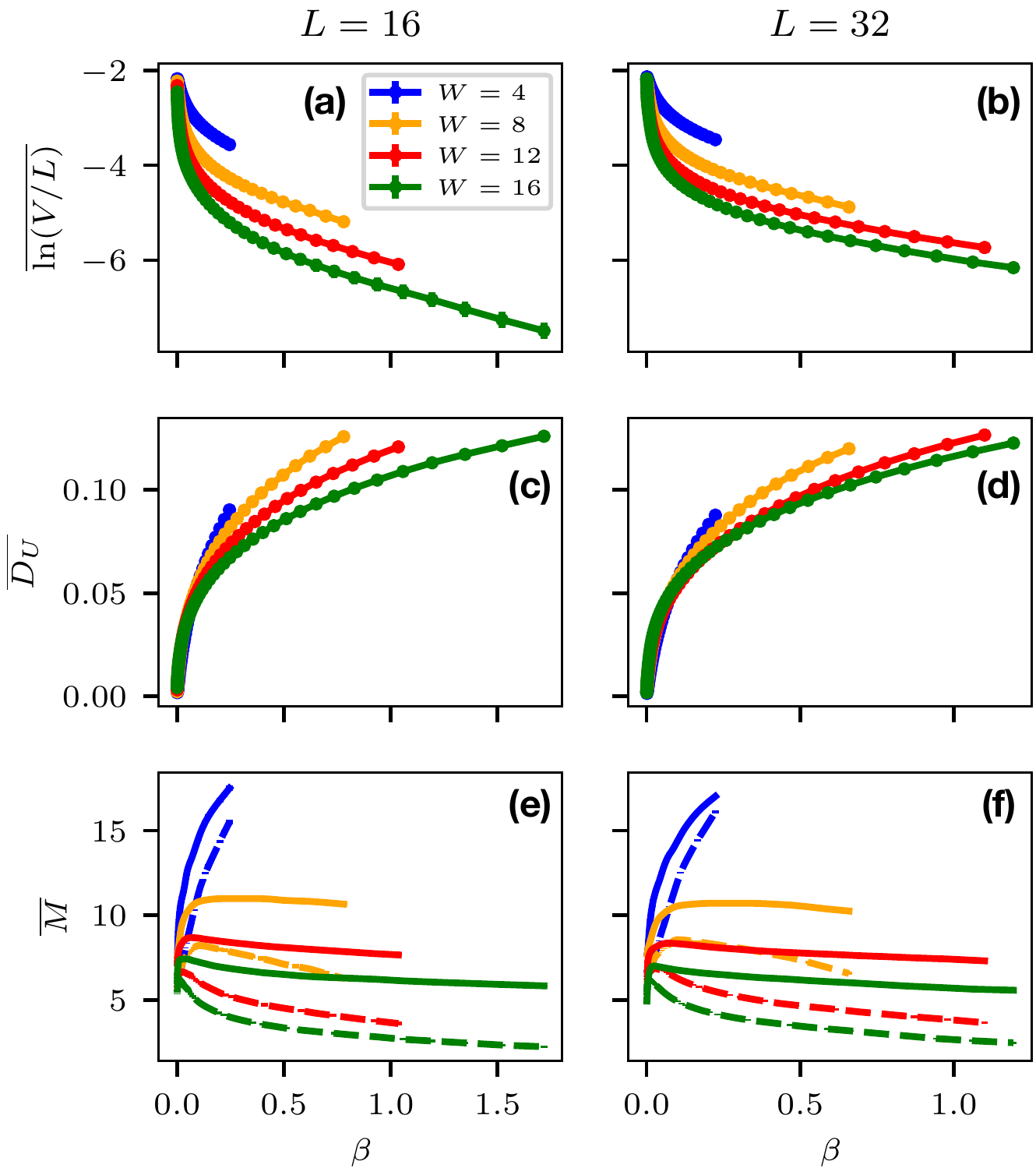}
\caption{Ensemble averaged MPO Wegner flow data  for $L=16$ (left) and $L=32$ (right). SVD cutoffs are  $4 \times 10^{-12}$ and $2 \times 10^{-12}$, respectively.  Shown are the average variance per site (top), the unitary distance (middle) and the average bond dimension (bottom) of the effective Hamiltonian $H(\beta)$ (solid) and $dU(\beta)$ (dashed).
}
\label{fig:mpo_Wegner_flow}
\end{figure}

Our numerical algorithm constructs the UTN, one row at a time, where each row corresponding to a single step of the Wilson-Wegner flow represented as the matrix-product operator $dU(\beta)$ of low bond-dimension (see Fig.~\ref{fig:UTN}(a)). To avoid the need to directly work with exponentially large matrices,  $H_0(\beta)$, $H_1(\beta)$,  and $U(\beta)$ are all represented as matrix-product operators (MPO); see app.~\ref{app:MPO_and_MPS} for a review of MPO. Operator addition is implemented as a direct sum of tensors for each site and multiplication as a direct product. The complexity of our algorithm scales linearly in the maximum $\beta$ and polynomially as $M(\beta)^5$ in the bond-dimensions $M(\beta)$ of $H(\beta)$.  The scaling in $\beta$ could be exponentially improved if an implicit time-stepping method was used in lieu of the (primarily) fixed $\Delta \beta$ we are using  (see app.~\ref{app:TensifiedWegnerFlow} for details of our algorithm). In Fig.~\ref{fig:mpo_Wegner_flow}(e,f) we see that deep in the many-body localized phase the bond-dimension of both $dU(\beta)$ and $M(\beta)$, at large $\beta$, is bounded by a constant with no noticeable system-size dependence.  App.~\ref{app:compare_cutoffs} validates the correctness of our approach.  Note that while our algorithm is designed for diagonalizing Hamiltonians, it will also disentangle a state $|\Psi\rangle$ if we let $H=|\Psi\rangle\langle \Psi|$. 

In Fig.~\ref{fig:mpo_Wegner_flow}(a,b) we measure the variance (equivalently average off-diagonal term) of the UTN
\begin{equation} \label{eq:avg_var}
V(\beta) \equiv \frac{1}{N} \left( \text{Tr}[H(\beta)^2]
- \sum_i E_i^2 \right)  \equiv \frac{1}{N} \sum_{i\neq j} H^2_{ij}(\beta),
\end{equation}
where we have assumed that $H(\beta)$ is real, $N=\dim(H)$,  $E_i= \langle i| U  H   U^\dagger | i \rangle$, and $|i\rangle$ is the i'th product state (see app.~\ref{app:Evaluate_V} for details on how the variance can be computed efficiently). We find that $V(\beta)$ decreases exponentially with $\beta$ which, combined with our bound on the bond-dimension, ensures the efficiency of TWWF.

While the induced tensor network is naively a grid (fig.~\ref{fig:UTN}(a)), there is significant variance in the auxiliary (horizontal) bond-dimension as a function of both site and $\beta$ leading to an emergent geometry in the bulk. In fact it is common (see fig.~Fig.~\ref{fig:UTN}(c)) to see large regions where the auxiliary bond-dimension is one indicating that $dU(\beta)$'s have decomposed into independent unitaries.  While the bond-dimension occasionally increases again later in the flow (see the bar at large $\beta$ in Fig.~\ref{fig:UTN}(c)) , this is at a much lower energy scale. To a reasonable approximation, we can view this initial loss of bond-dimension as signaling the primary disentangling of an l-bit.  We interpret these large-$\beta$ bars as resonances, transformations which span, but don't rotate the intermediate sites.  

We can see the action of the large-$\beta$ bars explicitly by considering the rate of change of the coupling constants $J_z^i(\beta) = Tr(\sigma^z_i H(\beta))$ of $H(\beta)$ (see Fig.~\ref{fig:spaghetti}). Notice that pairs of $dJ_i^z/d\beta$ tend to be anti-correlated (i.e. the red and purple lines in Fig.~\ref{fig:spaghetti}). This behavior is indicative of the Wilson-Wegner flow
working on the off-diagonal matrix element that connects sites $i$ and $j$. These terms become diagonalized when $\beta (h_i-h_j)^2 \sim 1$ where $h_i$ and $h_j$ are the corresponding (renormalized) single-site energies. On closer inspection, we observe higher order anti-correlations (e.g. the triplet ) which correspond to the Wilson-Wegner flow of higher order terms.  The RG time $\beta$ at which the bond-dimension of $dU(\beta)$ becomes order unity and $dJ/d\beta$ becomes small are roughly the same. 

We now consider the light-cone like spread of local operators induced by the UTN.  We find that in the MBL phase local operators initially spread as $r(\beta) \sim \log(\beta/\beta_0)$, where  $\beta_0$ is a constant setting the scale of $\beta$ and $r(\beta)$ is the ``horizontal'' size of the operator at a given $\beta$ (see Fig.~\ref{fig:lightCone}).  Once the operator hits the position-dependent ceiling (blue line Fig.~\ref{fig:UTN}c) the spreading stops. On the other hand, in the ergodic phase operators spread as $r(\beta) \sim \sqrt{\beta}$ (see Fig.~\ref{fig:lightCone}), and the spreading continues to the edge of the system. We note that the behavior of operator spreading under WWF is similar to the spread of entanglement under real-time evolution for MBL systems~\cite{Znidaric2008, Bardarson2012,Huse2014,Kim2014,Deng2017} and Lieb-Robinson bound for ergodic systems~\cite{Lieb1972}. 
The WWF sets a scale between energy and RG time, $\beta \propto 1/E^2$.  Therefore, in the MBL phase l-bit couplings decay exponentially $E(L) \propto \exp[-L/(2 L_0)]$ (till they reach the site-dependent ceiling) while in the ergodic phase they decay algebraically as $E(L) \propto 1/L$. In the MBL phase, this relation sets a natural length-scale $L_0$ for the decay of $\ell-$bits interactions.  See app.~\ref{app:collidingLightCones} for a similar analysis using colliding light cones. 

In addition to considering the spread of light-cones, we can also consider distances through the bulk.  To consider these distances, it is useful to understand the relationship between the unitary tensor network and the tensor network which generates eigenstates $|\Psi_i\rangle= U^\dagger |i\rangle$.  These tensor networks are identical except the latter is terminated at the top ($\beta=\infty$) of the UTN by a binary $\ell$-bit configuration $|i\rangle$ (i.e. a product state in the $S_z$ basis).  After such a termination, the tensor network can be additionally simplified by `pulling' to smaller $\beta$ the terminated legs at large $\beta$ which span sites over a bar (see fig.~1(b)). This transforms the UTN into a more MERA-like object where sites are decimated at smaller RG time.  This further motivates the idea that unit auxiliary bond-dimension should be considered `empty' in the bulk (as in fig.~1(c)) as there are then no vertical nor horizontal bonds in this region.

One natural way to think about distances in the bulk~\cite{Pastawski2015Holographic} is to consider the sum of the logarithm of bond-dimensions through a given cut -- i.e. the tensor cut distance.   In the vertical direction, this distance is sensitive to the time-step we use in our RG flow (although in app.~\ref{app:TensorDistance} we show evidence that the qualitative physics is largely insensitive to this). This motivates us to instead focus on an alternative vertical distance measure. We define the unitary distance $D_U$, over our UTN, as
\begin{equation}
D_U(\beta) = \int_0^{\beta} \sqrt{\frac{ Tr(\eta(\tau)\eta^{\dagger}(\tau)) }{\textrm{dim}(H)L}} \D\tau
\end{equation}
where $e^{\eta(\tau)\D\tau}$ is the infinitesimal unitary transformation at RG time $\tau$, and the factor of $1/\textrm{dim}(H)$ is included to rescale the trace of an identity operator in the many-body basis to 1. This distance generalizes the notion of the Burr metric used in cMERA~\cite{Nozaki2012,Mollabashi2014} to unitaries (see appendix~\ref{sec:BuresAppendix}). In cMERA this distance measures the rate of change of the quantum state with RG flow. In the WWF, the unitary distance is directly related to the rate at which the variance of the Hamiltonian shrinks.

To make explicit this relationship, we use WWF
\begin{align}\label{eq:decay}
\frac{dV}{d\beta} = \frac{1}{\text{dim}(H)} \frac{\D}{\D \beta} \text{Tr}(H_1^2) = -\frac{2}{\text{dim}(H)} \text{Tr} (\eta^{\dagger} \eta) \le 0,
\end{align} 
to obtain
\begin{equation}
D_U(\beta) = \int_0^\beta \sqrt{-\frac{1}{2L}\frac{dV(\tau)}{d\tau} } d\tau.
\end{equation}
In fig.~4(c) \& (d) we show the ensemble average of $D_U$ as a function of $\beta$.

While this gives us a notion of vertical distance, we still need to specify circumferential distance. We will use the the number of auxiliary bonds which are not unit bond-dimension.  Earlier we saw this was equivalent to the number of $J_i^z$ operators still being rotated in $H(\beta)$ and effectively measures the number of $\ell-$bits in the system which have not yet been diagonalized. It is also equivalent, when considering the eigenstate-version of the tensor network, to the number of vertical bonds cut by a path.  

\begin{figure}[th]
\centering
\includegraphics[width=\columnwidth]{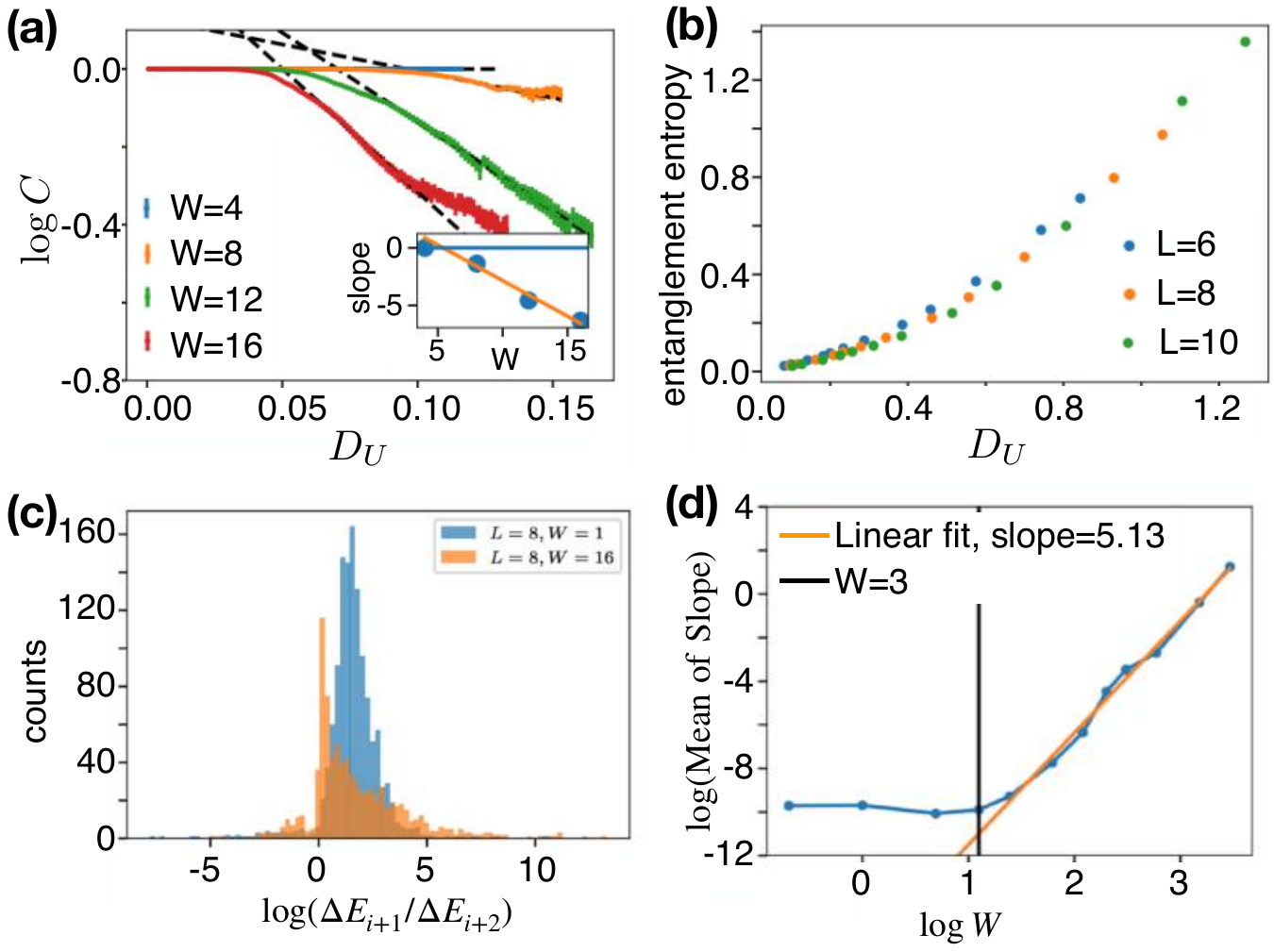}
\caption{\textbf{(a)} Ensemble averaged value of the logarithm of the circumference versus the unitary distance $D_U$ for $L=32$.  The inset shows the slope of these lines as a functions of $W$ (blue dots).  The orange line is the best fit to these points. 
\textbf{(b)} Relationship between final unitary distance and average entanglement entropy of eigenstates at various values of $W \in \{0.5,1,2,3,4,6,8,10,12,16,24,32\}$ in both the ergodic and MBL phase.   
\textbf{(c)} Histogram of the logarithm of the ratios of consecutive slopes of $\log V(\beta)$.  These slopes correspond to the energy level at which the Wegner-Flow is currently diagonalizing.  \textbf{(d)} Log-Log plot of the slope of $d\log \overline{V} / d\beta$ as $\beta\rightarrow \infty$ for $L=8$ as a function of $W$.
}
\label{fig:Circumference}
\end{figure}

We find that the circumference decays exponentially with vertical distance (see fig.~\ref{fig:Circumference}(a) and fig.~\ref{fig:L16Circumference}); using an alternative vertical metric (the tensor distance) gives qualitatively similar conclusions (see app.~\ref{app:TensorDistance}).  An exponentially decaying circumference is consistent with seeing minimally rare regions of all scales as such a rare region of length $r$ should appear with probability $\exp(-r)$ and have a vertical bulk distance of $r$ (corresponding to a volume law entanglement).  As the ergodic phase is approached, the coefficient of the exponential continuously approaches zero suggesting the rare regions percolate the system.  

 We find that the final $D_U$ is quadratically related to the half-cut bipartite entanglement averaged over eigenstates (see fig.~\ref{fig:Circumference}(b)). Surprisingly, the same quadratic relation holds for various chain lengths and disorder strength.  This relation is reminiscent of the RT formula~\cite{Ryu} that states that there is a correspondence between the entanglement at the boundary and the minimal geodesic in the bulk. In our case the final unitary distance $D_U$ is a proxy for the length of the minimal geodesic that appears when we cut the system in half to measure the bi-partite entanglement entropy.
 %Using the final unitary distance $D_U$ as a proxy for these geodesics, we can empirically test this relationship. 

Finally, while we have focused on the local real-space picture of the MBL phase, the Wegner flow is really an energy-based RG which probes different energy scales at different RG time.  As the variance decays as $V(\beta) \propto \exp[-\beta (\Delta E)^2]$ we can determine the energy scale at a given $\beta$ by looking for linear segments of $\log(V(\beta))$.  This is accomplished using a top-down linear segmentation method adapted from the Ramer-Douglas-Peucker (RDP) algorithm.  This is done only on small systems because of the need to reach $\beta \rightarrow \infty$.  In the ergodic phase, consecutive energy scales drop at a fixed exponential rate while in the MBL phase consecutive energy scales decrease monotonically at a rate which is (on average) exponential but whose distribution is significantly broadened (see fig.~\ref{fig:Circumference}(c)). Most interestingly, the final renormalized energy scale should be diagonalizing the lowest energy scale of the system (this scale can be generated from the large $\beta$ slope of $\overline{\log V/L}$).  In the ergodic phase, this is  at the interlevel spacing.  On the other hand, in the MBL phase, we find that the lowest energy scale is $\Delta E \sim W^{\alpha(L))}$ where $\alpha(L)$ is size-dependent (see fig.~\ref{fig:Circumference}(d)); this comes from the fact that energies which differ by the interlevel spacing aren't coupled after renormalization. These results imply that the geometry of the bulk at large $\beta$ is different in the two phases. Further, comparing where the ergodic and MBL curves cross, we find the critical point $W_c^{\alpha(L)/L} = 2$.

\section*{Acknowledgment}
We acknowledge useful conversations with Bela Bauer, Vadim Oganesyan, Vipin Varma, Sung-Sik Lee, Tom Faulkner, and
Steve Shenker who introduced us to the concept of shredded horizons.  We acknowledge support by SciDAC-DOE grant DE-FG02-12ER46875 (BKC and XY) as well as DOE DE-SC0020165 (BKC) during part of this work. DP acknowledges support from the Charles E. Kaufman foundation and NSF PIRE-1743717.  We thank the Aspen Center for Physics (DP and BKC) and the Galileo Galilei Institute (BKC) for their hospitality.
This research is part of the Blue Waters sustained petascale computing project, which is supported by the National Science Foundation (awards OCI-0725070 and ACI-1238993) and the State of Illinois. Blue Waters is a joint effort of the University of Illinois at Urbana Champaign and its National Center for Supercomputing Applications. 

\bibliography{QuantumBlock}

\clearpage
\appendix
\renewcommand\thefigure{S\arabic{figure}}    
\setcounter{figure}{0}    

\section{MPO and MPS}\label{app:MPO_and_MPS}
To describe the spin-$\frac{1}{2}$ Hamiltonians and the local unitary transformations, we will use the matrix product operator (MPO) representation, where an operator $A$ with support on $n$ consecutive sites will be represented as
\begin{equation} \label{eq:MPO}
O = \sum_{\{\mathbf{\sigma}\},\{\mathbf{\sigma}'\}} A_1^{[\sigma_1,\sigma_1']} \cdots A_n^{[\sigma_n,\sigma_n']} | \sigma_1 \cdots \sigma_n \rangle \langle \sigma_1' \cdots \sigma_n' |.
\end{equation}
For any choices of site index $i$ and spin  indices $\sigma_i$ and $\sigma_i'$, $A_i^{[\sigma_i,\sigma_i']}$ is a matrix of size $M \times M$, except at the edges where the tensors are actually vectors. $M$ is usually called the bond dimension. On each site, there are 4 different matrices, $A_i^{[+,+]}$, $A_i^{[+,-]}$, $A_i^{[-,+]}$ and $A_i^{[-,-]}$.

\begin{figure}[th]
\centering
\includegraphics[width=0.26\textwidth]{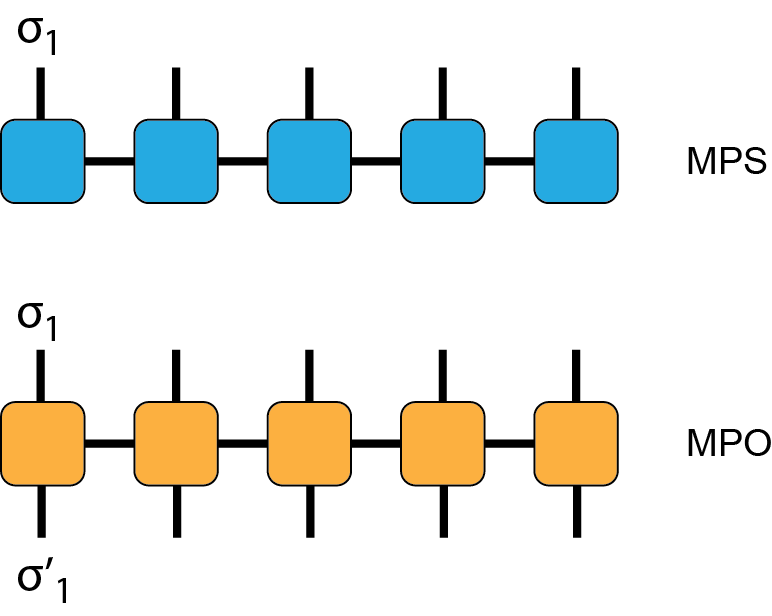}
\caption{Graphical illustration of a MPS and a MPO.}
\label{fig:mps_mpo}
\end{figure}

If the diagonal part of $A$ is needed as an independent MPO, it can be easily seen that 
\begin{equation} \label{eq:diag_MPO}
O_{\text{diag}} = \sum_{\{\mathbf{\sigma}\}} A_1^{[\sigma_1,\sigma_1]} \cdots A_n^{[\sigma_n,\sigma_n]} | \sigma_1 \cdots \sigma_n \rangle \langle \sigma_1 \cdots \sigma_n |.
\end{equation}
To generate this MPO, we simply fill the matrices $A_i^{[+,-]}$, $A_i^{[-,+]}$ with zeros for each site, thereby effectively dropping the off-diagonal terms from $A$.

Similarly, a wave-function $|\psi\rangle$ for a $n$ site spin-$\frac{1}{2}$ system can be represented as a matrix product state (MPS)
\begin{equation} \label{eq:MPS}
|\psi\rangle = \sum_{\{\mathbf{\sigma}\}} A_1^{\sigma_1} A_2^{\sigma_2} \cdots A_n^{\sigma_n} | \sigma_1 \sigma_2 \cdots \sigma_n \rangle,
\end{equation}
where $A_i^{\sigma_i}$ is again a matrix.

More information on the operations of MPSs and MPOs can be found in ref.~\onlinecite{Schollwock2011}.

\section{Algorithm for Tensor Wilson-Wegner Flow}\label{app:TensifiedWegnerFlow}

At small system sizes, one can afford to numerically integrate the flow equations using a sparse matrix format~\cite{Pekker2016Fixed}. For arbitrary disorder strength, we can use the Runge-Kutta Fehlberg 4(5) method with dynamical time step to perform the integration until the average variance of the Hamiltonian drops close to machine precision.

At larger system sizes, the sparse matrix representation is generally impractical. Below we describe the tensor Wilson-Wegner flow algorithm which uses matrix product operators (MPO).  We apply the following steps:
\begin{enumerate}
\item The generator $\eta(\beta)$ can be rewritten as
\begin{equation}
\eta(\beta) = [H_0(\beta), H(\beta)],
\end{equation}
where the diagonal MPO $H_0(\beta)$ can be constructed easily by dropping the off-diagonal matrices on each site of $H(\beta)$'s MPO.  (See appendix \ref{app:MPO_and_MPS}).

\item We time evolve the flow equations, eqn.~\eqref{eq:Flow_Equation}, using an explicit method. During a small, finite time step $\Delta \beta$, the unitary operator becomes
\begin{equation}
U(\beta + \Delta \beta) = dU(\beta)U(\beta).
\end{equation}
where $dU(\beta) \equiv \exp(\eta(\beta) \Delta \beta)$
While the simplest approach to obtaining $dU$ is to use a Taylor series approximation to the  exponentiation\cite{stoudenmire2010minimally} and evolve the Hamiltonian $H(\beta)$ according to
\begin{equation}
H(\beta + \Delta \beta) = e^{\eta(\beta) \Delta \beta} H(\beta)  e^{-\eta(\beta) \Delta \beta},
\end{equation}
we find it more stable and accurate to expand and approximate $H(\beta + \Delta \beta)$ directly using the Baker-Campbell-Hausdorff (BCH) formula.
\begin{multline} \label{eq:mpo_wegner_H}
H(\beta +\Delta \beta)   =  \ H + \Delta \beta [\eta, H] +\\
 \frac{\Delta \beta^2}{2!} [\eta, [\eta, H]] +  \frac{\Delta \beta^3}{3!} [\eta, [\eta, [\eta, H]]] + \cdots \\
 = \ H + \Delta \beta \Bigg[\eta, \ H +  \\ \frac{\Delta \beta}{2}  \bigg[\eta, \ H + \frac{\Delta \beta}{3} \Big[ \eta,\ H + \cdots \Big] \bigg] \Bigg]
\end{multline}

The slight change of form in the last equation is important allowing us to repeatedly evaluate terms like $H + \frac{\Delta \beta}{n} [\eta, \ H']$, whose bond dimensions are much better controlled compared to the exponentially (with respect to the number of $\eta$) growing bond dimensions  of $[\eta, [\eta, [...,[\eta, H]...]]$.
\end{enumerate}

During the above procedure, we usually use a fixed SVD cutoff. It is important to notice that time evolutions of $H(\beta)$ and $U(\beta)$ can be carried out independently, and the former is possible without even building the $U(\beta)$.

\section{Efficacy of Tensor Wilson-Wegner Flow} \label{app:compare_cutoffs}
\begin{figure}[th]
\centering
\includegraphics[]{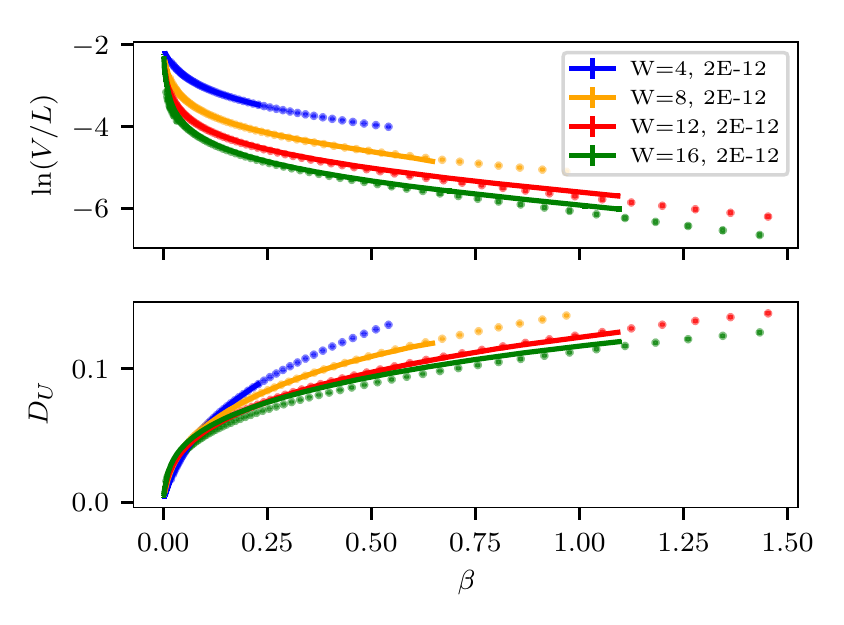}
\caption{(Colored online) Comparison of ensemble averages of $\overline{\ln (V/L)}$ between SVD cutoffs of $2 \times 10^{-10}$ (dots) and $2 \times 10^{-12}$ (lines), at $L=32$ and with the same 100 disorder realizations.  The simulations are not converged at this $\beta$ but run for a finite wall-clock time. It can be seen that there is no significant difference between the two sets of curves.
}
\label{fig:compare_svd_cutoffs}
\end{figure}

In this appendix, we consider the efficacy of TWWF.  To begin with, when running our algorithm, we compress our MPO at each step using a SVD cutoff.  Here we report the effect of this SVD cutoff in fig~\ref{fig:compare_svd_cutoffs} comparing the ensemble average of $\ln (V/L)$ and $D_U$ for $L=32$ using SVD cutoffs of $2 \times 10^{-12}$ and $2 \times 10^{-10}$.

\begin{figure}[th]
\centering
\includegraphics[width=0.42\textwidth]{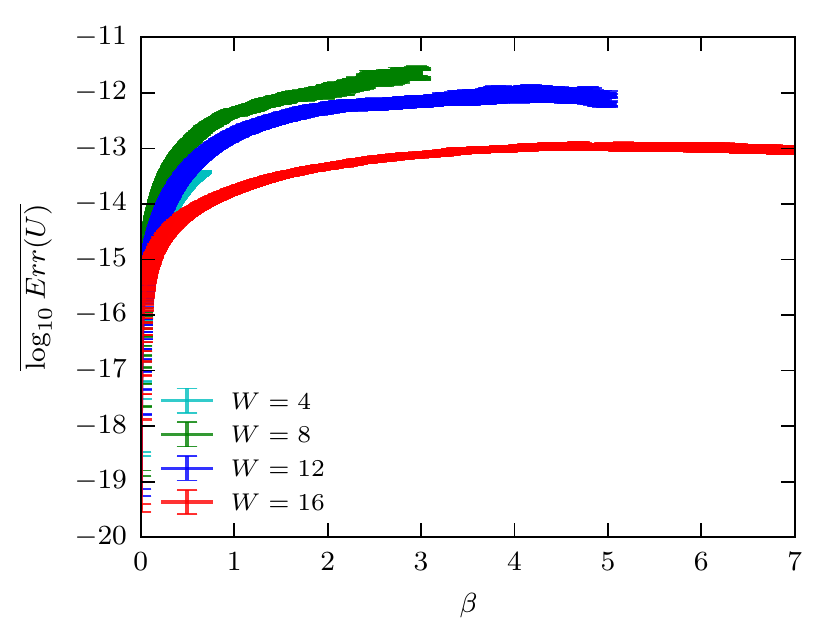}
\caption{ \label{fig:Unitarity} Disorder-averaged errors on the unitary operators $U(\beta)$ during the Wegner flow in Fig. \ref{fig:mpo_Wegner_flow} with 100 disorder realizations for each disorder strength $W$ at $L=32$. The calculations are done with a SVD threshold of $2\times 10^{-11}$. The errors are defined as $\text{Err}(U) = |UU^{\dagger}-\mathbb{I}|^2/\text{dim}^2(H)$, which essentially measure the difference per matrix element between $UU^{\dagger}$ and $\mathbb{I}$. We can see that at large $\beta$, the errors are well controlled and do not seem to diverge.}
\end{figure}

We can also measure the errors produced by the effect of truncation by verifying that we have not seriously broken unitarity.  To check this we consider the error per element in the unitary matrix in fig.~\ref{fig:Unitarity} as a function of $\beta$.  Note that this quality is adjustable by tuning the SVD cutoff $\epsilon.$

\begin{figure}[th]
\centering
\includegraphics[width=0.42\textwidth]{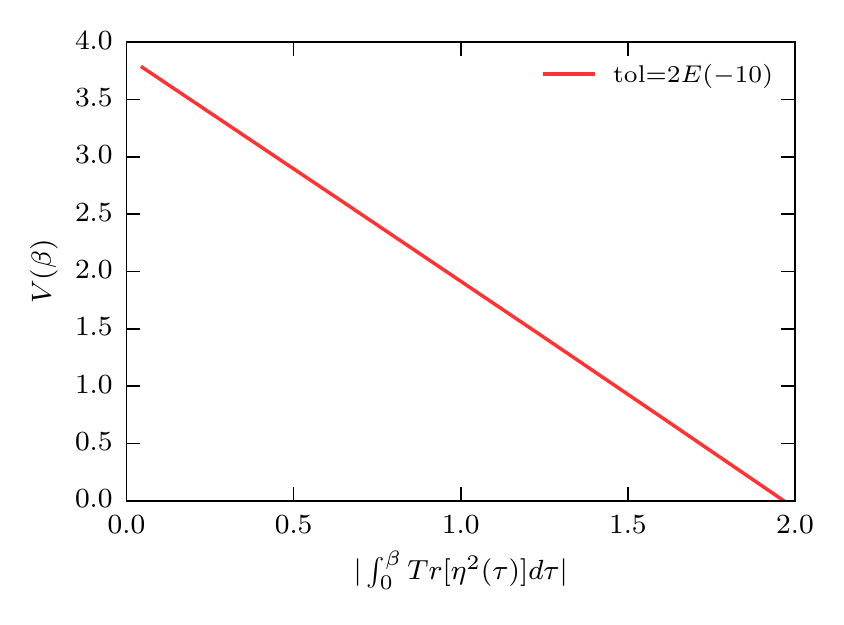}
\caption{Average variance $V$ of a disorder sample with $L=32,W=16$, plotted against the integral of $|\text{Tr}(\eta^2)|$ with respect to RG time. The calculation is performed using the MPO representation, with SVD truncation error bounded by $2\times10^{-10}$. A constant factor of $1/\text{dim}(H)$ is absorbed into trace operation.}
\label{fig:var_vs_etasq}
\end{figure}
Finally we know that during Wegner flow, since the trace and the $l^2$-norm of the total Hamiltonian $H(\beta)$ are invariant under the unitary transformation, it is easy to see that the change rate of the average variance  $V$ is proportional to the change rate of the $l^2$-norm of the off-diagonal Hamiltonian $H_1(\beta)$. So, from Eq. \ref{eq:decay} we have
\begin{equation}
\frac{\D}{\D \beta} V = -2 \frac{1}{\text{dim}(H)} \text{Tr} (\eta^{\dagger}(\beta) \eta(\beta)).
\end{equation}
This linear relation is shown in Fig. \ref{fig:var_vs_etasq}, which further verifies the accuracy and validity of our MPO implementation of the Wegner flow.

\section{Algorithm to evaluate average variance} \label{app:Evaluate_V}
When using the MPO version of the Wegner flow, we need an efficient method of evaluating the average variance $V$ defined in Eq. \eqref{eq:avg_var}. For spin-$\frac{1}{2}$ systems, $V$ can be rewritten as
\begin{equation}
V(\beta) \equiv \frac{1}{2^L} \sum_{\{\tau\}} \left[ \langle \{\tau\}| H^2(\beta) |\{\tau\} \rangle - \langle \{\tau\}| H(\beta) |\{\tau\} \rangle^2 \right].
\end{equation}
where $L$ is the system size, $H(\beta)$ is the Hamiltonian at the RG time $\beta$, and $\tau$ represent a product state defined in the $\tau_z$ basis (approximate l-bit basis).

Although the above expression involves a summation over exponentially many states, one can easily avoid exponential time cost, using alternative interpretation of the terms in the MPO language.

The first part of the summation, in the MPO notations, can be written as
\begin{equation}
\sum_{\{\tau\}}  \langle \{\tau\}| H^2 |\{\tau\} \rangle = \sum_{\{\tau\},\{\tau'\}} \left( A_1^{[\tau_1,\tau'_1]} A_2^{[\tau_2,\tau'_2]} \cdots A_L^{[\tau_L,\tau'_L]} \right)^2.
\end{equation}

Similarly, the second part of the summation can be written as
\begin{equation}
\sum_{\{\tau\}}  \langle \{\tau\}| H |\{\tau\} \rangle^2 = \sum_{\{\tau\}} \left( A_1^{[\tau_1,\tau_1]} A_2^{[\tau_2,\tau_2]} \cdots A_L^{[\tau_L,\tau_L]} \right)^2,
\end{equation}
which is simply the L2 norm of the  diagonal MPO $H_{\text{diag}}$. We obtain $H_{\text{diag}}$, as discussed after eqn.~\eqref{eq:diag_MPO}.

Both parts can be efficiently evaluated using canonicalization techniques at a cost of $O(2LM^3)$, where $L$ is the system size and $M$ is the typical bond dimension.

\section{Radial distance and average Bures distance}
\label{sec:BuresAppendix}
In this appendix we review the quantum information motivation for using the unitary distance $D_U$ as our metric. In quantum information, given two states described by density matrices $\rho_1$ and $\rho_2$, the Bures distance $D_B$ between them is defined through
\begin{equation}
D_B^2 = 2(1 - Tr \sqrt{ \rho_1^{1/2} \rho_2 \rho_1^{1/2} }).
\end{equation}
For pure states $\rho_1 = |\psi_1\rangle \langle \psi_1 |$ and $\rho_2 = |\psi_2\rangle \langle \psi_2 |$, one gets
\begin{equation}
D_B^2 = 2(1 - \langle \psi_1 | \psi_2 \rangle ).
\end{equation}
Observe that the $D_B^2$ defined here depends on the system size $L$. 

Inspired by Ref. \onlinecite{Nozaki2012,Mollabashi2014}, we defined the radial metric of the unitary tensor network generated by Wilson-Wegner flow as 
\begin{equation}
g_{\tau\tau} d\tau^2 = \frac{2}{\text{dim}(H)L} Tr(\mathbb{I} - e^{\eta d\tau} ) = \frac{Tr(\eta^{\dagger}\eta)}{\text{dim}(H)L} d\tau^2,
\end{equation}
where $\eta$ is the anti-Hermitian generator and $L$ is the system size. The metric defined above is essentially the infinitesimal per unit length Bures distance brought by the unitary transformation of $e^{-\eta d\tau}$, averaged over a complete set of pure states. We include a factor of $1/L$ to remove some system size dependence from the radial distance of the RG flow.

The radial distance from the boundary (p-bits) to a RG time $\beta$ in the bulk of the unitary tensor network is given by
\begin{equation}
D_U(\beta) = \int_0^{\beta} \sqrt{g_{\tau\tau}} d\tau = \int_0^{\beta} \sqrt{ \frac{Tr(\eta^{\dagger}\eta)}{\text{dim}(H)L} } d\tau.
\end{equation}

\section{Tensor Distance}\label{app:TensorDistance}
\begin{figure}[th]
\centering
\includegraphics[width=0.4\textwidth]{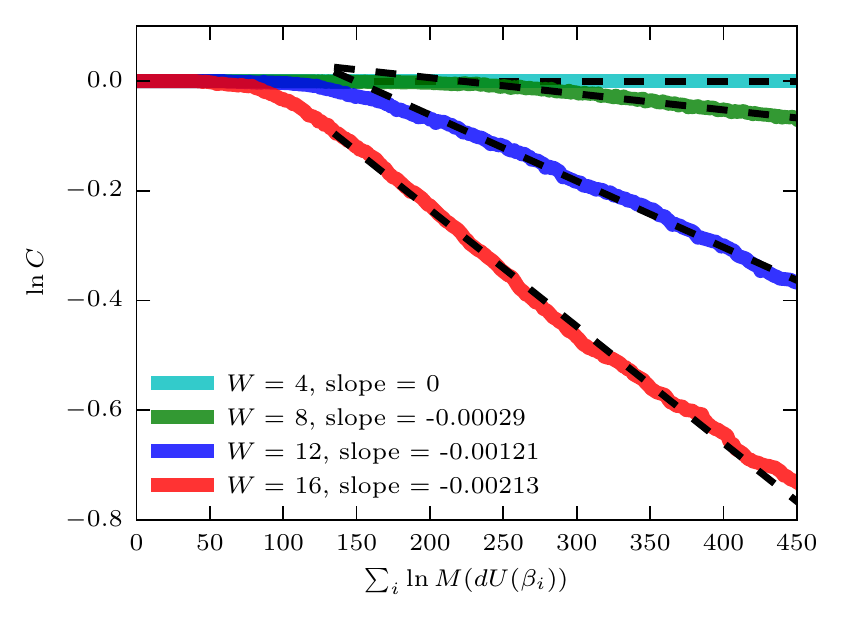}
\includegraphics[width=0.4\textwidth]{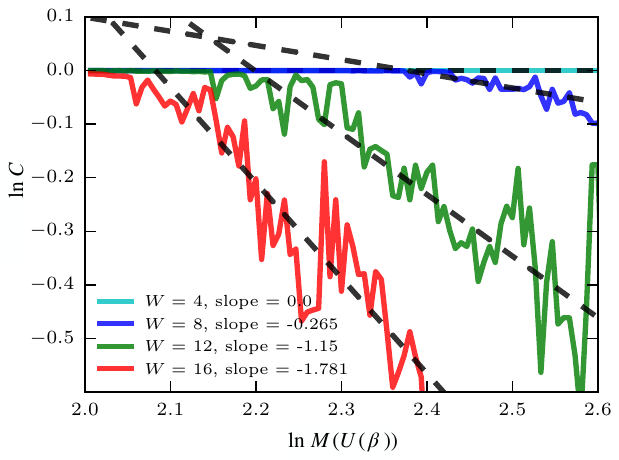}
\includegraphics[width=0.4\textwidth]{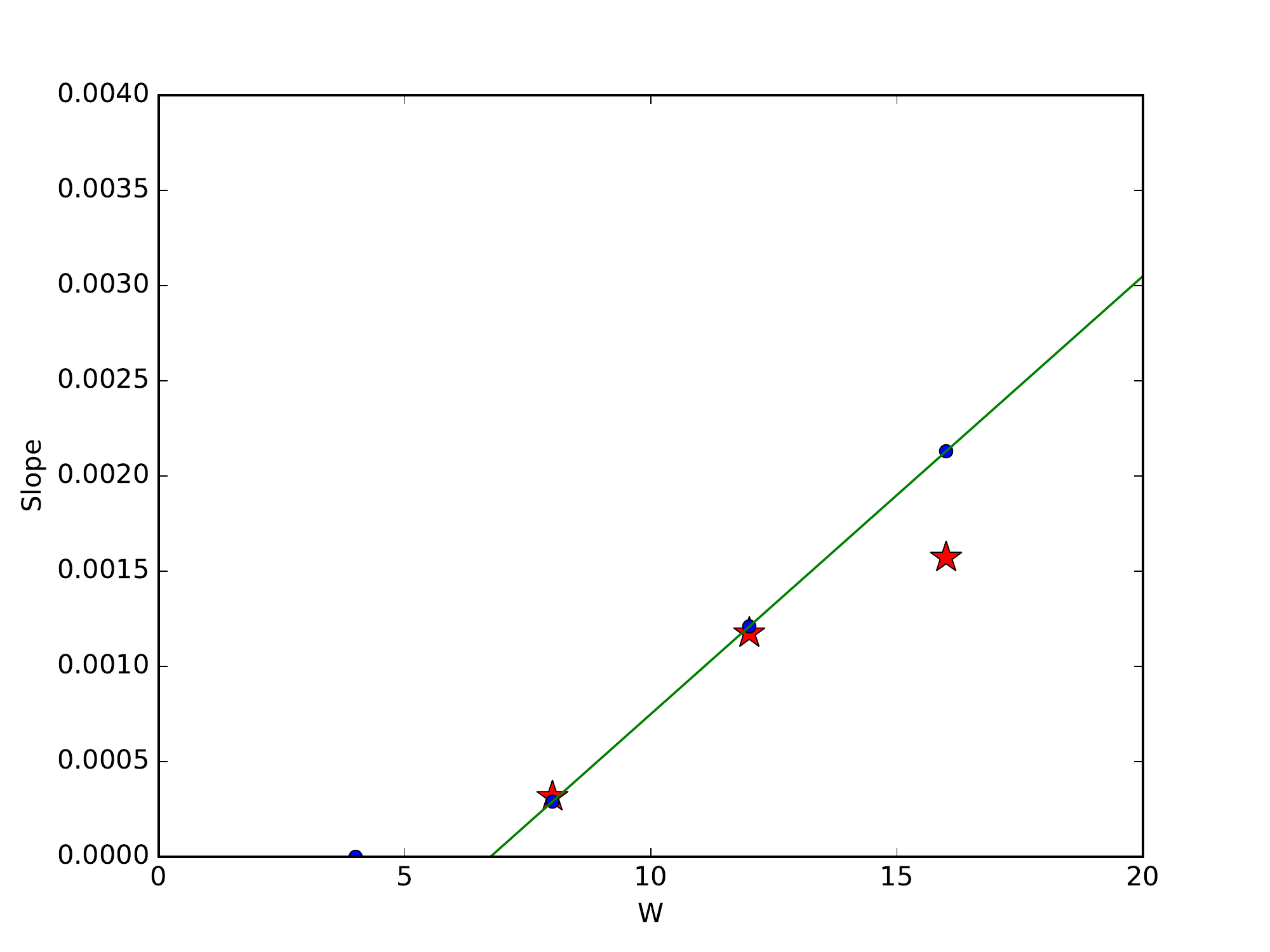}
\caption{Logarithm of circumference vs. tensor distance for $L=32$ using bond-dimensions of individual $dU$ (\textbf{top}) and the entire UTN compressed to a single $M$ (\textbf{middle}).  
 \textbf{Bottom:} Dots are slope of exponential decay as a function of disorder strength $W.$ Stars indicate same fit done for distance generated from $M(U)$ scaled by 1089.}
\label{fig:CircVsDistance}
\end{figure}

In the main text, we have focused on using the unitary distance $U_D$.  While this distance is well motivated for the Wilson-Wegner flow, another common distance to use is the tensor distance which is the sum of the logarithm of the bond-dimensions
\begin{equation}
T(\beta)=\sum_i \ln M(dU(\beta_i))
\end{equation}
where $M(\cdot)$ is the bond dimension of a tensor and $i$ indexes over all the sites. 
We show the results for the tensor distance in fig.~\ref{fig:CircVsDistance}(top).  The tensor distance is sensitive to the choice of $\tau$ which sets the number of MPO which stack to get to a given $\beta$.  To understand this sensitivity, we considered the other extreme where we compress the UTN into a single tensor $M$ (see fig.~\ref{fig:CircVsDistance}(middle)).  While this significantly changes the values of the slopes (on the semi-log plot), the ratios of these slopes are similar (see fig.~\ref{fig:CircVsDistance}(bottom)).  

\section{Colliding Light Cones}\label{app:collidingLightCones}
In the main text, we measured the rate at which light cones spread.  Here we take an alternative approach measuring instead the RG time $\beta$ at which two operators evolved under unitary evolution take to collide.  We find that the light cones collide at a $\beta$ where their initial separation $L=\log(\beta)$ out to some cutoff distance consistent with the logarithmic light cone spread found in the text. 

\begin{figure}[th]
\centering
\includegraphics[width=\columnwidth]{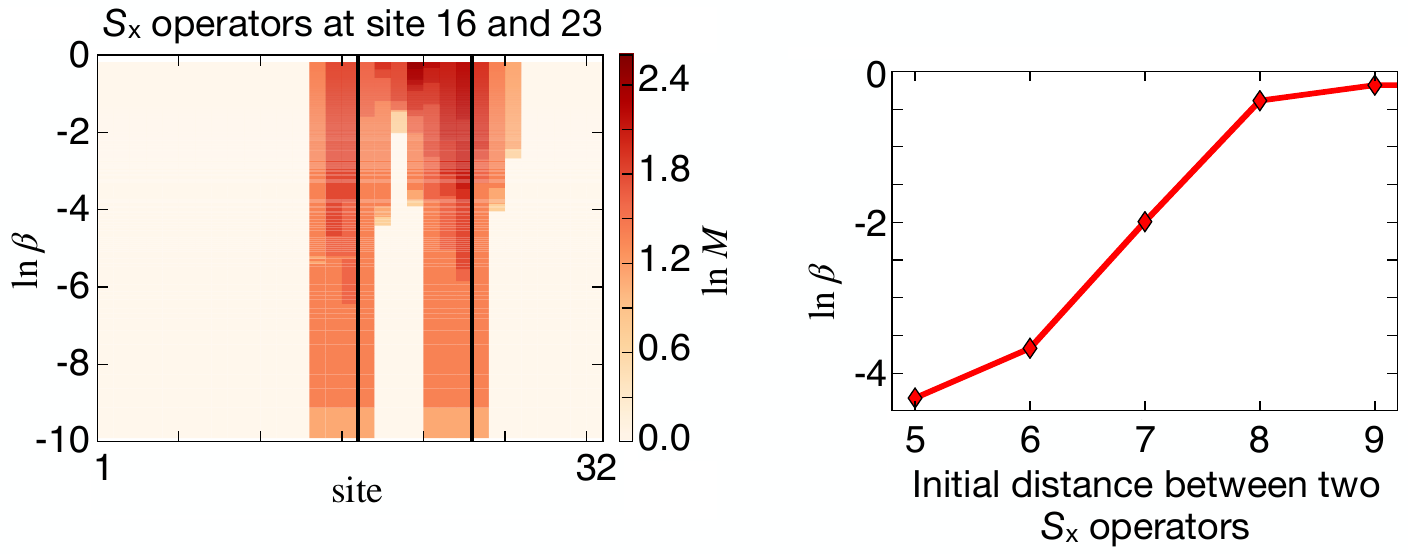}
\caption{\textbf{Left: }Bond dimension of $F(\beta)$ for a prototypical configuration at $L=32; W=16$.  \textbf{Right: } $\beta$ at which two operators a distance $d$ away take to collide.}
\label{fig:collidingLightCones}
\end{figure}
We checked this explicitly in the following way.  On a prototypical configuration deep in the MBL phase, we set two operators $\sigma^x$ at sites $k$ and 23 and evolve them as $F(\beta) \equiv U(\beta) \sigma_{k}^x \sigma_{23}^x U^\dagger (\beta)$ (see fig.~\ref{fig:collidingLightCones}(left)).  We  consider the light cones to have ``collided'' when the bond-dimension of the resulting operator becomes greater then one between sites 23 and $k$.  At small $d=|23-k|$ they collide at the smallest accessible $\beta$ and at large $d$ they never collide having reached the diagonal state before they would intercept.  At intermediate $d$ though we find that they scale as $\log \beta$ (see fig.~\ref{fig:compare_svd_cutoffs}(right)).  

\section{Removing l-bits}
\begin{figure}
\centering
\includegraphics[width=\columnwidth]{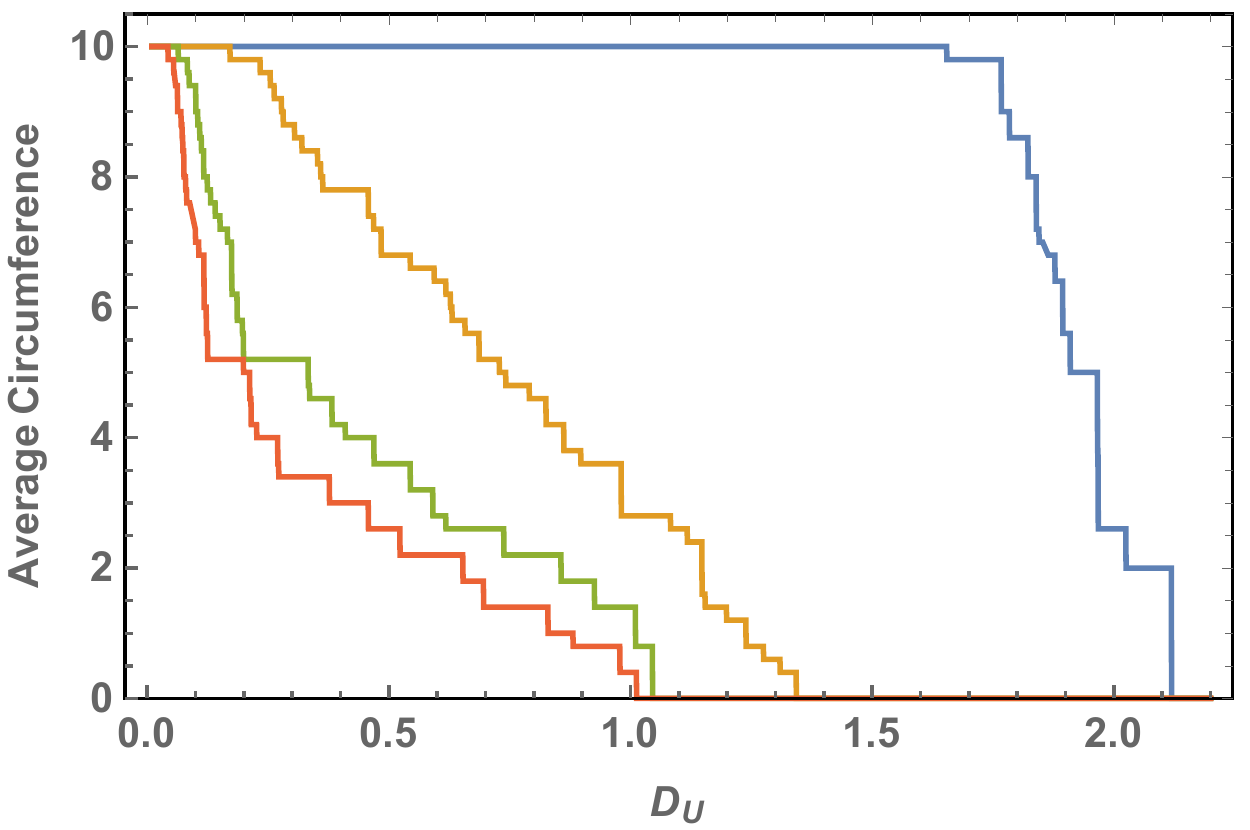}
\caption{
Rate at which l-bits are diagonalized from the system for MBL phase (red $W=12$, green $W=8$), the transition region (yellow $W=3.5$) and the ergodic phase (blue $W=1$) at $L=10$. 
}
\label{fig:s8}
\end{figure}

In the main text, we considered the rate at which l-bits were diagonalized in the RG flow by measuring the circumference as a function of $D_u$. We were able to consider this for large systems ($L\approx 32$) in the MBL phase using  TWWF.  Here, for smaller systems, we consider a similar analysis using ED Wegner flow but for disorder strengths spanning the entire range from localized to ergodic through the transition region. We still find that in the MBL phase, the l-bits are diagonalized at a rate which is consistent with being exponential with system size.  On the other hand, in the ergodic region, we find that the l-bits are all diagonalized only at large $D_u$.  
Interestingly, in the transition region, we find that l-bits are diagonalized at a rate which is uniform in $D_u$  (see fig.~\ref{fig:s8}

\section{Additional Figures}
\begin{figure}[th]
\centering
\includegraphics[scale=0.7]{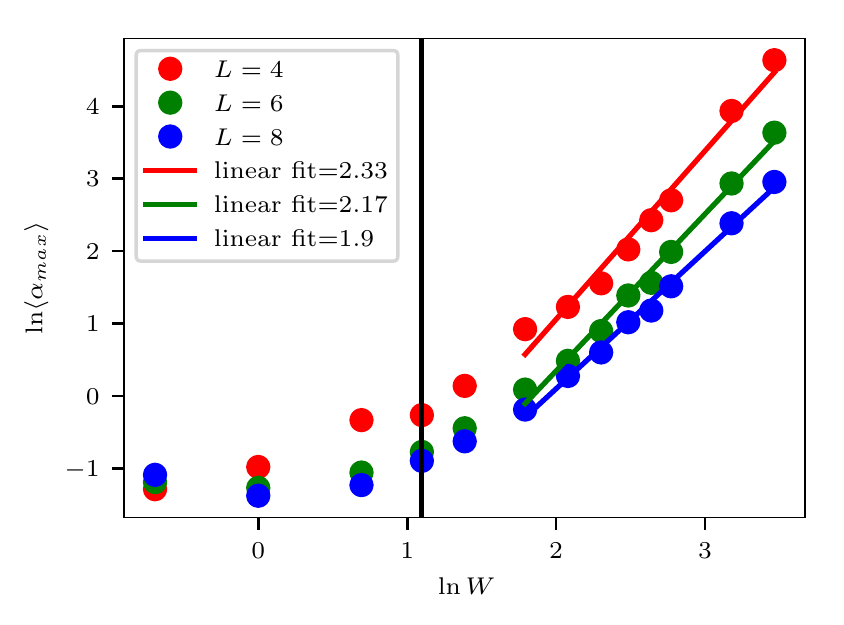}
\includegraphics[scale=0.7]{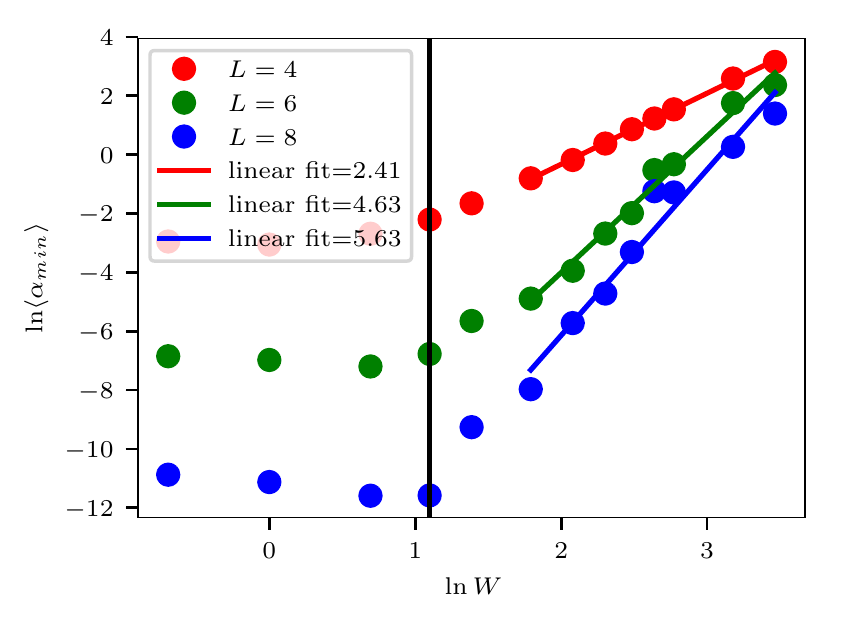}
\caption{Average over 200 disorder realizations of  first identifiable slope $\alpha_\textrm{max}$ (left) and last identifiable slope $\alpha_\textrm{min}$ (right) of $\overline{\ln(V/L)}$  }
\label{fig:lbits}
\end{figure}

In this section we include some additional graphs which supplement the information in the main text.  In fig.~\ref{fig:lbits} we see the first and last slope of $\overline{\ln(V/L)}$  identified by the Ramer-Douglas-Peucker (RDP) algorithm.  We expect the slopes to go las $\Delta E^2$ for the current energy scale $E$; as the largest energy scale goes as $W$, we see that the largest slopes go roughly as $W^2$.  On the other hand, the last slopes are similar to those seen in fig.~\ref{fig:Circumference}(bottom right).  

\begin{figure}[th]
\centering
\includegraphics[width=0.4\textwidth]{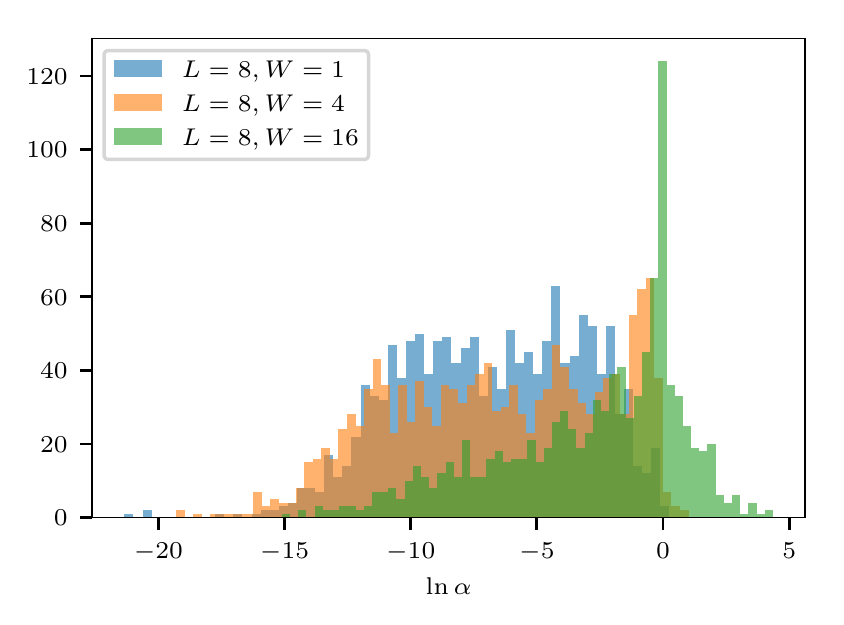}
\caption{
Histogram of the distribution of $\ln \alpha$ (log of slopes), for $L=8$ and $W=\{1,4,16\}$ of 200 disorder realizations. These slopes are those used to produce the ratios in fig.~\ref{fig:Circumference}
}
\label{fig:alpha}
\end{figure}

In fig.~\ref{fig:alpha} we show the distribution of slopes of $\ln\alpha$ for various values of $W$.  

\begin{figure}[th]
\includegraphics[width=0.95\columnwidth]{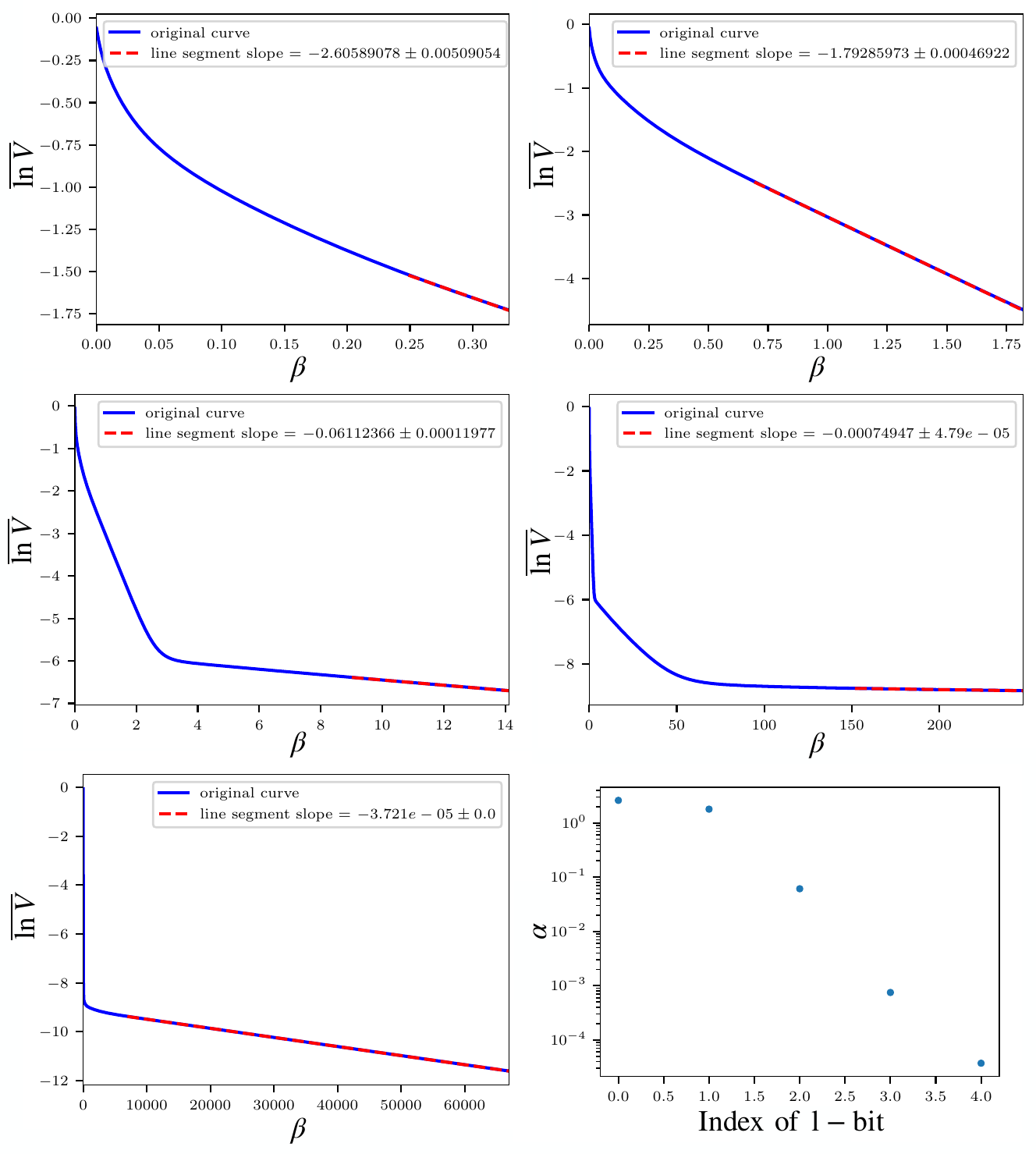}
\caption{
Decay of the first five slopes of a sample for $L=8, W=12$ for various ranges of $\beta$ with line segments identified by the Ramer-Douglas-Peucker(RDP) algorithm. Final graph shows these slopes. This is meant to be an illustrative but not necessarily  typical sample. Final graph is slopes versus order found. }
\label{fig:slopes}
\end{figure}

\begin{figure}[th]
\centering
\includegraphics[width=0.4\textwidth]{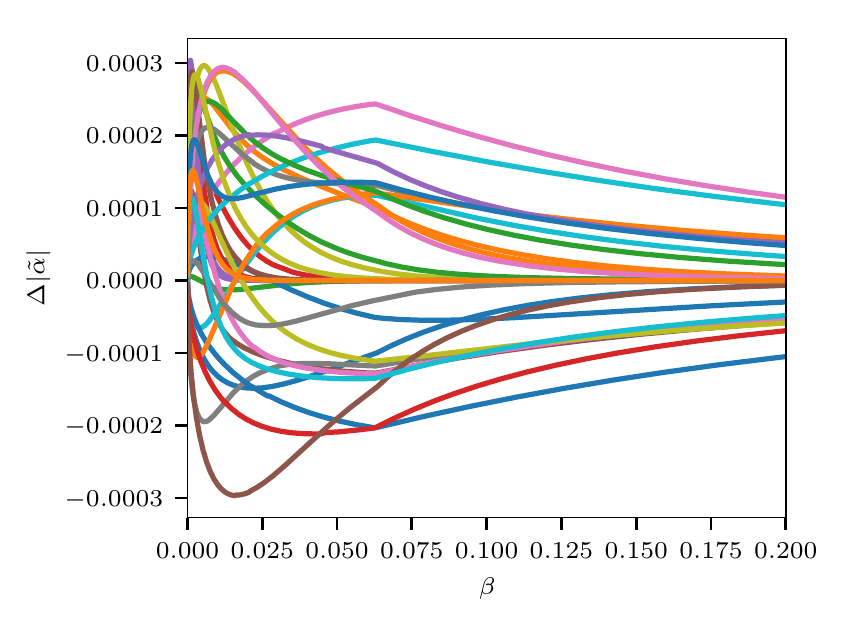}
\includegraphics[width=0.4\textwidth]{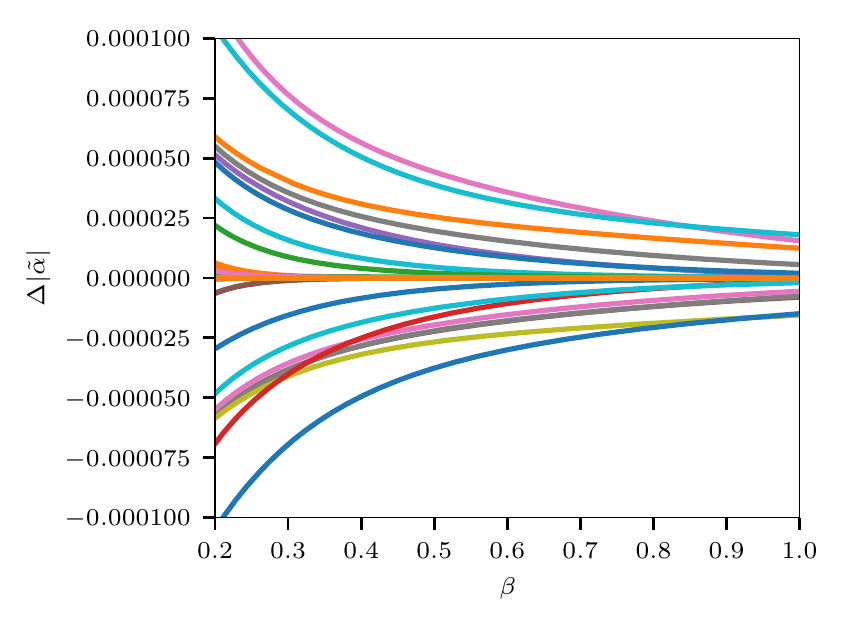}
\caption{
Rate of change of $dJ_i^z/d\beta$ for all sites $i$ for the sample in fig.~\ref{fig:spaghetti} at smaller $\beta$. 
}
\label{fig:allSpaghetti}
\end{figure}

\begin{figure}[th]
\centering
\includegraphics[]{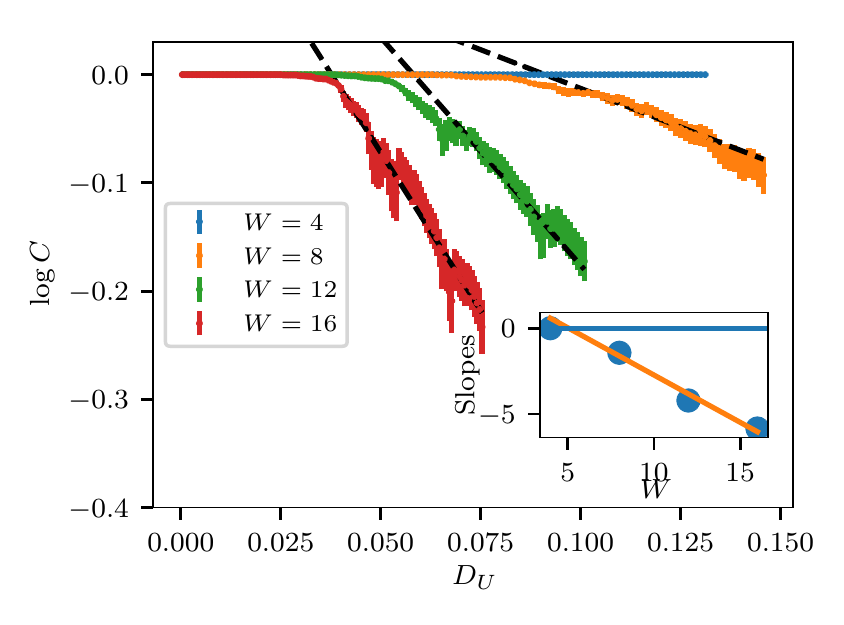}
\caption{Ensemble averaged value of the logarithm of the circumference versus the unitary distance $D_U$ for $L=16$
}
\label{fig:L16Circumference}
\end{figure}

In fig.~\ref{fig:slopes} we show an illustrative sample of the slopes identified by the RDP algorithm, in fig.~\ref{fig:allSpaghetti} we see the rate of change of the l-bit coupling constants at more $\beta$ then displayed in the main text,  and in fig.~\ref{fig:L16Circumference} we see the exponential decrease of the circumference for $L=16$

\end{document}